\documentclass[preprint,showpacs,preprintnumbers,amsmath,amssymb]{elsarticle}

 \usepackage{graphics}
 \usepackage{array}

\usepackage{amsmath}
 \usepackage{amsthm}

 \usepackage{lineno}
\usepackage{tikz}
\usepackage{graphics}

\begin{document}
\begin{frontmatter}



\title{Providing physics guides in Bayesian neural networks from input layer: case of giant dipole resonance predictions}
%  Injecting physical information
\author[1]{Xiaohang Wang}
\author[1]{Jun Su \corref{cor2}}
\author[1]{Long Zhu}
\address[1]{Sino-French Institute of Nuclear Engineering and Technology, Sun Yat-sen University, Zhuhai 519082, China}

\cortext[cor2]{Corresponding author. sujun3@mail.sysu.edu.cn}

\begin{abstract}
  The Bayesian neural network (BNN) has been applied to evaluate and predict the nuclear data.
  However, how to provide physics guides in BNN is a key but an open question.
  In this work, the case study on giant dipole resonance (GDR) energy is presented to illustrate the effectiveness and maneuverability of the method to provide physics guides in BNN from input layer.
  The Spearman's correlation coefficients are applied to assess the statistical dependence between nuclear properties in the ground state and the GDR energies.
  Then the optimal ground-state properties are employed as the input layer in the BNN for evaluating and predicting the GDR energies.
  Those selected ground-state properties actively contributes to reduce the predicted errors and avoid the risk of the non-physics divergence.
  This work gives a demonstration to find effects of the GDR energy by using the BNN without the physics motivated model, which may be helpful for discovering physics effects from the complex nuclear data.
  % In this contribution
  % A minimal error or deviation may result in wide divergence.失之毫厘，谬以千里。
  % to account for the physics that is missing from the most sophisticated macroscopic models.
  % to incorporate as much physics as possible

\end{abstract}


\end{frontmatter}

% \linenumbers

%\section{}
%\label{}\
%\section{\label{int}Introduction}
Bayesian machine learning is a powerful tool to make predictions with confidence interval after training by data.
In this respect, the Bayesian neural network (BNN) has rapidly drawn much attention in nuclear physics in the past few years.
How to provide physics guides in the BNN approach, which is merely a numerical algorithm, is a key problem when applied in the physics field.
The seminal ideas on this topic can date back to the early works using neural networks in nuclear physics \cite{Gernoth1993, Athanassopoulos2004, Costiris2009, Bayram2014}.
A clear illustration of the underlying philosophy behind the implementation of the BNN approach originally appears in works by Utama et al., where the cases of theoretical predictions of nuclear masses \cite{Utama2016} and nuclear charge radii \cite{Utama2016a} were applied to validate that one can include as much physics as possible in the initial prediction by using the physics motivated models and use the BNN to fine tune these models by modeling the residuals.
To facilitate the narrative, this approach is hereinafter called residual-approach.

The residual-approach has been proved to be a very powerful tool to improve the physical model-based predictions of, such as, nuclear binding energies \cite{Neufcourt2018}, fission yields \cite{Wang2019}, and isotopic cross-sections in spallation \cite{Ma2020}.
Beyond the residual-approach, successful attempts to further consider the physical information into the BNN approach can also be found in the literature.
For example, it was found that better predictive performance for nuclear masses can be achieved by adding two quantities related to the well known nuclear pairing and shell effects into the input layer of the BNN \cite{Niu2018}.
However, as stated in Ref. \cite{Utama2016}, the basic requirement of the residual-approach is the existence of a robust theoretical model to provide physics guides.
Without the physics motivated model, can one find physics effects from the complex big data by using the BNN?
This work is an attempt to answer this open question through the case study on the prediction of giant dipole resonance (GDR) energy.

The GDR is the most well known and oldest collective excited mode of nucleus in the excitation energy from single nucleon separation energy to dozens of MeV \cite{Baldwin1947, Ishkhanov2007, Bracco2019}.
Over the last decades, hundreds of GDR data have been experimentally measured using photonuclear reactions or $\gamma$ decay \cite{Ceruti2017, Mondal2018, Kobayashi2019}.
The $\gamma$ spectrum displays a wide peak, which can be characterized by a Lorentzian function with GDR parameters including the peak energy (i.e. GDR energy), the width and the strength \cite{Tian2019, Plujko2018}.
Various models have been developed to investigate the GDR, including Hartree-Fock-based random phase approximation (HF-based RPA) \cite{Lyutorovich2015, Bonasera2018}, time-dependent Hartree-Fock theory (TDHF) \cite{Fracasso2012, Mennana2020}, Boltzmann-Uehling-Uhlenbeck model \cite{Wang2020, Kong2017}, extended quantum molecular dynamics \cite{He2014}, and macroscopic Langevin equation coupling with the Skyrme Hartree-Fock-Bogolyubov model \cite{Su2019, Su2020}.
%The photonuclear cross sections in the GDR region are significant in many engineering fields such as nuclear energy and medical applications \cite{Capote2009}.
%The GDR parameters are also excellent probes in many topics in nuclear physics, because of their strong relations to the microscopic structure of the nuclei.
Several effects of the GDR, such as the mass dependence, the isospin dependence, and the shape effect, have been studied \cite{Colo2014,Su2019, Wang2020, Chakrabarty2016, Pandit2019}.
Those effects were also found in the nuclear properties in the ground state, the data of which are more abundant than those of GDR \cite{Huang2017, Audi2017, Wang2017}.
In fact, proposing by the microscopic theory such as HF-based RPA and TDHF, the GDR energies depend strongly on the HF ground-state wave function \cite{Bonasera2018, Mennana2020}.
In the macroscopic point of view, the GDR energy has been determined when the proton and neutron density distribution in the ground state is known \cite{Su2019}.
Thus, there may exist dependence between the ground-state properties and the GDR energies.

\begin{figure*}[htb]
	\centering
	\includegraphics[width=16.0cm]{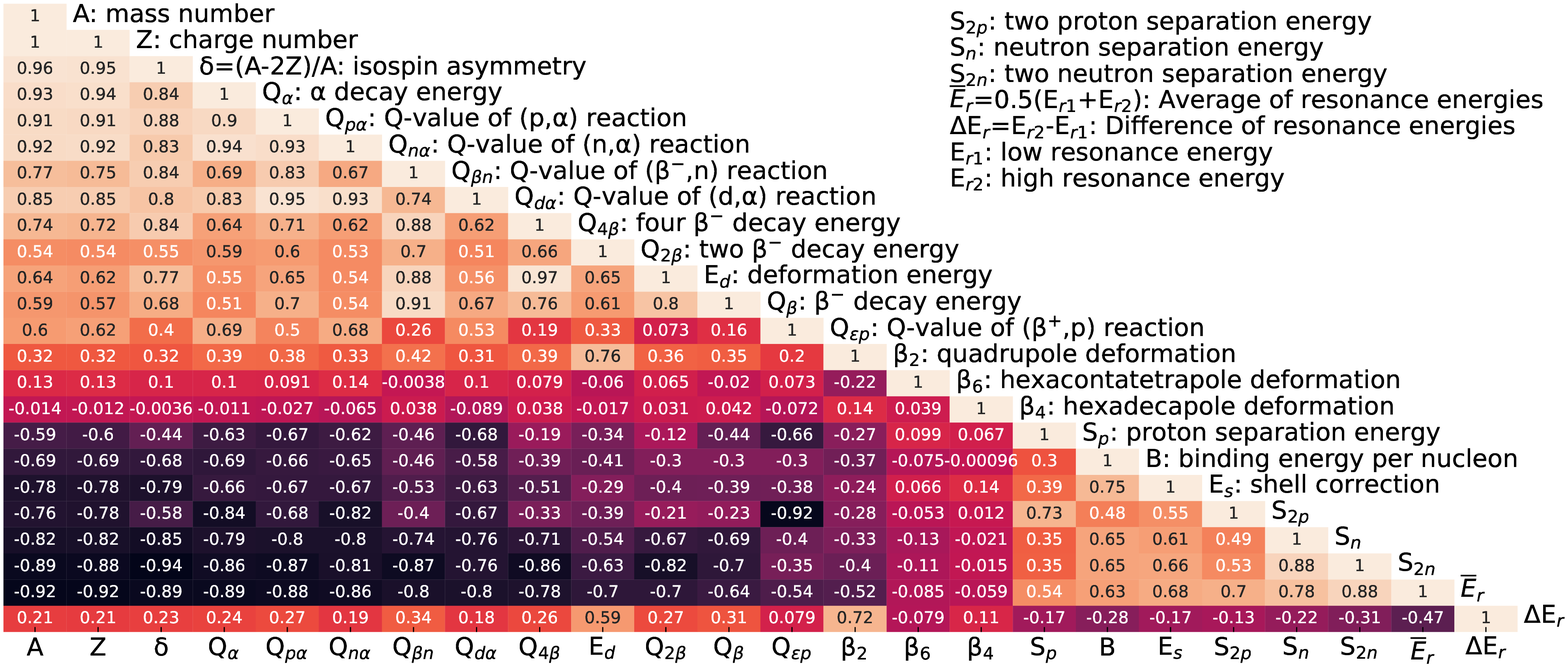}
	\caption{Spearman's correlation coefficient $\rho$ between ground-state properties and GDR energies. The pairs of variables will be strongly positive correlated for high $\rho$ value (or identical for a correlation $\rho$ = 1), and strongly negative correlated for a low $\rho$ value (or fully opposed for a correlation $\rho$ = -1).
A coefficient $\rho$ near zero indicates that there is no correlation between pairs of variables.
The ground-state properties are taken from \cite{Huang2017, Audi2017, Wang2017, Wang2014}.
The data of the GDR energy are taken from Ref. \cite{Plujko2018}.}
	\label{fig-1}
\end{figure*}

To assess the statistical dependence between two variables of the ground-state properties and GDR energies, the Spearman's correlation coefficients $\rho$(x,y) are calculated,
\begin{equation}
\rho(x,y) = \frac {\sum_{i}(x_{i}-\bar{x})(y_{i}-\bar{y})} {\sqrt{\sum_{i}(x_{i}-\bar{x})^{2}\sum_{i}(y_{i}-\bar{y})^{2}}},
\end{equation}
where $x_{i}$ is the i-th data of the variable x, $\bar{x}$ is the average of the variable x over the data, and $\sum_{i}$ expresses the summation over the data.
Similar situation is for the variable y.
As listed in Fig. \ref{fig-1}, the ground-state properties include the derived quantities of atomic masses taken from the Ame2012 evaluation \cite{Huang2017, Audi2017, Wang2017}, and the extracted quantities from the Weizs\"{a}cker-Skyrme mass formula \cite{Wang2014}.
The GDR energy include the low and high peak energies (E$_{r1}$ and E$_{r2}$), which are taken from Ref. \cite{Plujko2018}.
Generally speaking, there are two components of the GDR energy for quadrupole deformation, but is only one for spherical nuclei.
In Ref. \cite{Plujko2018}, the high resonance energies are missing for spherical nuclei.
We let E$_{r2}$ = E$_{r1}$ when high resonance energy is missing.
It means that two peaks of the GDR spectrum overlap for spherical nuclei.

Intuitively, The pairs of variables will be strongly positive correlated for high $\rho$ value (or identical for a correlation $\rho$ = 1), and strongly negative correlated for a low $\rho$ value (or fully opposed for a correlation $\rho$ = -1).
Those two cases both means that those two variables contain similar physical information, and hence one variable can be used to predict another.
A coefficient $\rho$ near zero indicates that there is no correlation between pairs of variables.
The variables may contain independent physical information.
To find the ground-state properties strongly depended on the average of resonance energies, one reads $\rho(A, \overline{E}_{r}$) = $\rho(Z, \overline{E}_{r}$) = -0.92 from Fig. \ref{fig-1}.
As the strongly dependence between mass and charge numbers [$\rho(A, Z)$ = 1], only one of them can be applied for the prediction of the resonance energies.
The strong isospin dependence of the GDR energies is also found, i.e. $\rho(\delta, \overline{E}_{r}$) = -0.89.
On the other hand, the dependence between the mass number A and the isospin asymmetry $\delta$ is strong [$\rho(A, \delta$) = 0.96].
Therefore, the isospin dependence of the GDR energies for the available data is covered by mass dependence.
For the difference of resonance energies $\Delta E_{r}$, correlation coefficients with the quadrupole deformation [$\rho(\Delta E_{r}, \beta_{2}$) = 0.72] and the deformation energy [$\rho(\Delta E_{r}, E_{d}$) = 0.59] are much larger than others.
It is indicated that the splitting of the GDR energies is a quadrupole deformation effect.
%But the quadrupole deformation $\beta_{2}$ does not include all physical information of the splitting.
For the dependence between the ground-state properties, one can find not only the strongly dependent pairs, such as $\rho(Q_{p\alpha}, Q_{d\alpha}$) = 0.94, but also the independent pairs, such as $\rho(Q_{\varepsilon p}, Q_{2\beta}$) = 0.073.
Those four quantities, i.e. $Q_{p\alpha}$, $Q_{d\alpha}$, $Q_{\varepsilon p}$, and $Q_{2\beta}$, are all derived from the data of the binding energy per nucleon.
It is indicated that various derivations of the atomic mass data put redundant information into the data set, but results in more independent physical information.
When those data set are used to predicte the GDR energies, useful information is handpicked by the BNN approach.

%\section{Bayesian Neural Network}
Early works of BNN include Refs. \cite{MacKay1992, bishop1995neural, neal2012bayesian}, which laid the foundation in this area.
Here it is introduced briefly.
The mapping from ground-state properties $X$ (input layer) to the GDR energies $E_{r}$ (output layer) is established by the neural network with one hidden layer.
\begin{equation}
E_{r}(X, \theta) = a+\sum_{j=1}^{H}b_j\mbox{logsig}\left(c_j+\sum_{i=1}^{l}d_{ji}X_i\right),
\end{equation}
where $\theta = \{a, b_j, c_j, d_{ji}\}$ are the parameters in the neural network, $\{a, c_j\}$ are biases, and $\{b_j, d_{ji}\}$ are weights.
The sigmoid function $\mbox{logsig}$ is used as the activation function.
$H$ is the the number of hidden neurons, and $l$ is the number of the input neurons.
The outputs are two-dimensional variable $E_{r} = \{E_{r1}, E_{r2}\}$.
The inputs $X = \{X_1,X_2,\ldots, X_l\}$ include one or several ground-state properties shown in Fig. \ref{fig-1}.

Using the Bayesian probabilistic model, the parameters $\theta$ in the neural network are expanded to random variables with distributions.
The posterior distribution of parameters $\theta$ given data set D is expressed as ,
\begin{equation}
P(\theta|D) = \frac{P(D|\theta)P(\theta)}{\int P(D|\theta)P(\theta)d\theta},
\end{equation}
where $P(\theta)$ is the prior distribution of $\theta$, $P(D|\theta)$ is the likelihood of $D$ given $\theta$.
$D = \{X^{(n)},E_{r}^{(n)}\}_{n=1}^{N_{d}}$ include both the ground-state properties $X^{(n)}$ and the GDR energy $E_{r}^{(n)}$,
where $N_{d}$ is the number of the available data.
%, $n = 1,2,\cdots, N_{d}$

For a nucleus known ground-state properties $X^{*}$, the expected value of GDR energy $\hat{E}_{r}^{*}$ is expressed as the integration,
\begin{equation}
	\hat{E}_{r}^{*}= \int E_{r}(X^{*}, \theta)P(\theta|D)d\theta,
\end{equation}
The Monte Carlo techniques are relied on to calculate the above integration,
\begin{equation}
	\hat{E}_{r}^{*} \approx \frac{1}{N_{s}}\sum_{k}^{N_{s}}E_{r}(X^{*}, \theta^{(k)}),
\end{equation}
%=E_{P(\theta|D)}E_{r}(X^{*}, \theta)
%where $E_{P(\theta|D)}$ denotes the Monte Carlo sampling drawn from the posterior distributions $P(\theta|D)$,
where $\theta^{(k)}$ ($k=1,2,\ldots, N_{s}$) is the k-th sample drawn from the posterior distributions $P(\theta|D)$, and $N_{s}$ is the number of the samples.
The 95\% confidence level is used to assess the uncertainty of the prediction.
The confidence interval of the GDR energy $E_{r}^{*}$ is,
\begin{equation}
	E_{r}^{*} = \hat{E}_{r}^{*} \pm 1.96\frac{\sigma}{\sqrt{N_{s}}},
\end{equation}
where $\sigma$ is the standard deviation of the samples $E_{r}(X^{*}, \theta^{(k)})$.

As the computation of the posterior distribution $P(\theta|D)$ is intractable due to the high dimension of parameters, the variation inference \cite{Mullachery2018, salimans2015markov, Blundell2015} is applied to find an approximation of $P(\theta|D)$.
The variation inference tries to find $\kappa$ so that $q(\theta|\kappa)$ is of minimum distance from $P(\theta|D)$ measured by Kullback-Leibler (KL) divergence:
  \begin{equation}
  \begin{aligned}
\theta =&\arg\min\mbox{KL }[q(\theta|\kappa)||P(\theta|D)] \\
       =&\arg\min E_{q(\theta|\kappa)}\left[\log \frac{q(\theta|\kappa)}{P(\theta|D)}\right] \\
       =&\arg\min E_{q(\theta|\kappa)}\left[\log \frac{q(\theta|\kappa)P(D)}{P(D|\theta)P(\theta)}\right] \\
       =&\arg\min  \sum_{k}\big[\log q(\theta^{(k)}|\kappa) - \log P(\theta^{(k)}) \\
       &- \log P(D|\theta^{(k)})\big].
  \end{aligned}
  \end{equation}

445 sets of data are available.
One hidden layer with 10 neurons is employed.
The standard normal distribution is used as the prior of weights.
1000 iterations are considered for the variation inference.
10000 samples are drawn for each predicted variable.

\begin{figure}[htb]
	\centering
	\includegraphics[width=8.5cm]{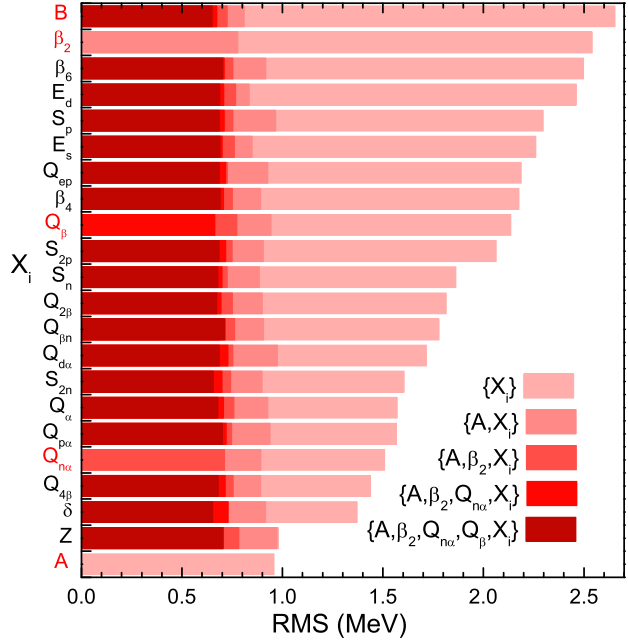}
	\caption{Root-mean-square deviations between data and BNN predictions. }
	\label{fig-2}
\end{figure}

The BNN predictions are evaluated by the root-mean-square (RMS) deviations, %evaluated 这个词用得对吗
  \begin{equation}
    RMS = \sqrt{\frac{1}{2N_{d}}\left[\sum^{N_{d}}_{n=1} (\hat{E}_{r1}^{(n)}-E_{r1}^{(n)})^{2} +\sum^{N_{d}}_{n=1}(\hat{E}_{r2}^{(n)}-E_{r2}^{(n)})^{2} \right]},
  \label{E1AI} %这个评估公式是否跟小航用得一样，主要看1/2n，如果不是，纵轴需要做缩放
  \end{equation}
where $E_{r1}^{(n)}$ and $E_{r2}^{(n)}$ are the n-th data of low and high GDR energies, $\hat{E}_{r1}^{(n)}$ and $\hat{E}_{r2}^{(n)}$ are their predicted values in the BNN.

One dimensional input variable $X = \{X_{i}\}$ is considered, where $X_{i}$ is one of the ground-state properties.
The RMS deviation for each input ground-state properties is shown as bar marked by \{X$_{i}\}$ in Fig. \ref{fig-2}.
It is found that the minimum of the RMS deviations is obtained when using input of mass or charge number (RMS = 0.96 or 0.98 MeV respectively).
This is consistent with the fact in Fig. \ref{fig-1} that the strongest correlation is found between the average of GDR energies and mass or charge number.
It is also consistent with the fact that the empirical formula aA$^{-1/3}$ is often used.
The further comparison between Spearman's correlation coefficient $\rho(X_{i}, \overline{E}_{r})$ and RMS deviations reveals their positive correlation.

According to the smallest RMS deviation, the mass number A is selected.
Then the input ground-state properties are expanded to two dimensional $X = \{A, X_{i}\}$.
Their RMS deviations are marked by $\{A, X_{i}\}$ in Fig. \ref{fig-2}.
It is found that adding one dimensional in the inputs generally improves the BNN predictions compared to $X = \{A\}$.
As one knows, the mass and charge numbers (or mass number and isospin asymmetry) are independent variables, which can be the identification of the nuclide.
Thus, in the residual-approach to study the nuclear charge radii \cite{Utama2016a}, fission yields \cite{Wang2019}, and isotopic cross-sections in spallation \cite{Ma2020}, the mass and charge numbers are applied as neurons of the input layer.
However, the results in Fig. \ref{fig-2} show that the RMS deviations for neither $\{A, Z\}$ nor $\{A, \delta\}$ are not the minimum among those for two dimensional input layer.
The reason is that the redundant physical information are included in data of mass and charge number.
In other words, the isospin dependence of the GDR energies for the available data is covered by mass dependence.
More data is of great necessity to reveal the isospin dependence.
For the available data, neither $\{A, Z\}$ nor $\{A, \delta\}$ is not the optimal input layer.

The minimum RMS deviation 0.78 MeV for two dimensional input is obtained by using $\{A, \beta_{2}\}$, which is nineteen percent lower than that using one dimensional input $\{A\}$.
It is consistent with the quadrupole deformation effect found from Fig. \ref{fig-1}.
Based on $\{A, \beta_{2}\}$, the input data are added to three, four, and five dimensions in sequence.
According to the minimum of the RMS deviation, the input variables $Q_{n\alpha}$, $Q_{\beta}$, and $B$ are handpicked.
The optimal ground-state properties to train the BNN for predicting GDR energies is $X = \{A, \beta_{2}, Q_{n\alpha}, Q_{\beta}, B \}$.
The RMS deviation is 0.65 MeV.

\begin{figure}[htb]
	\centering
	\includegraphics[width=8.0cm]{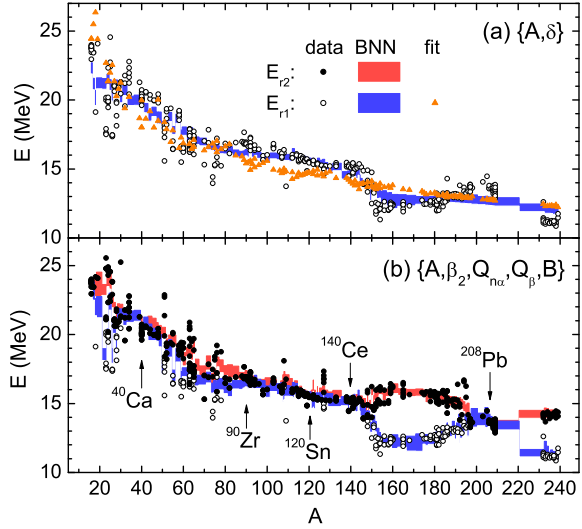}
	\caption{Data of GDR energies compared to the predictions by BNN using (a) two dimensional input layer $\{A, \delta\}$, and (b) five dimensional input layer $\{A, \beta_{2}, Q_{n\alpha}, Q_{\beta}, B \}$. The fitting by the empirical formula $(a+b\delta)A^{-1/3}$ is also shown as (orange) triangles.}
	\label{fig-3}
\end{figure}

The advantage of the BNN predictions using the optimal input $\{A, \beta_{2}, Q_{n\alpha}, Q_{\beta}, B \}$ comparing to those using two dimensional input $X = \{A, \delta\}$ can be found in Fig. \ref{fig-3}.
The 95\% confidence interval of predictions by the BNN approach with input $\{A, \delta\}$ is shown in Fig. \ref{fig-3}(a) as (blue) band.
The global decrease of the data with the increasing mass is reproduced.
However, the data in the regions near A = 150, 200, and 238 deviate from global mass dependence.
From A = 175 to 200, the value of $E_{r1}$ even increase with the increasing mass.
Those features of the data can not be reproduced by the BNN approach using input $\{A, \delta\}$.

The BNN predictions using the input $\{A, \beta_{2}, Q_{n\alpha}, Q_{\beta}, B \}$ are shown as (blue and red) bands in Fig. \ref{fig-3}(b).
In the figure, the arrows point out the cases of nuclei with magic number 20, 50, 82, and 126.
It is found that, for the nuclei near the shell, the two GDR energies are the same, which means there is only a peak in the GDR spectrum.
This feature is reproduced by the BNN predictions.
In the region out of the shell, there are differences between the low and high GDR energies.
Especially in the region 150 $<$ A $<$ 196 (60 $<$ Z $<$ 78 and 90 $<$ N $<$ 118), the differences between two GDR energies are nearly 4 MeV.
After trained by the data including the ground-state properties, the BNN predictions display the deformation effect.

It is worth to compare the predictions by the BNN and empirical formula with same input variables, i.e. $X = \{A, \delta\}$.
The empirical formula with parameters is often proposed by the physics motivated model or phenomenology.
For example, based on the mass dependence, the empirical formula aA$^{-1/3}$ is used to fit the data of the GDR energies.
In order to include the higher order effect, such as the isospin effect, the empirical formula is extended to $(a+b\delta)A^{-1/3}$.
By fitting the data of the low GDR energy $E_{r1}$, the parameters $a = 61.6 \pm 0.5$ and $b = 66.9 \pm 4.0$ are obtained.
As shown in Fig. \ref{fig-3}(a), the formula $(a+b\delta)A^{-1/3}$ reproduces the global decrease of the data with the increasing mass.
But the detailed features of the data in the regions near A = 150, 200, and 238 can not be reproduced.

The description and prediction of the empirical formula may be further improved by adding terms for the higher order effects, such as the shell effect or deformation effect.
Nevertheless, it's not always straightforward as several prerequisites are needed.
What is the next important effect?
How the effect can be expressed?
Is there any coupling with the know effects?
In fact, GDR energies depend on the ground-state properties, the data of which are abundant and
credible.
The BNN approach provides a method for mining useful information from the data of ground-state properties without the above prerequisites and ordering the importance of the effects, as shown in Fig. \ref{fig-2}.
With this importance-order, the effects may be further studied by the physics motivated model.
That is to say, the physics motivated model is significant both for the empirical formula and the BNN approach.
It is used aforehand for the empirical formula but afterward for the BNN approach.

\begin{figure}[htb]
	\centering
	\includegraphics[width=8.0cm]{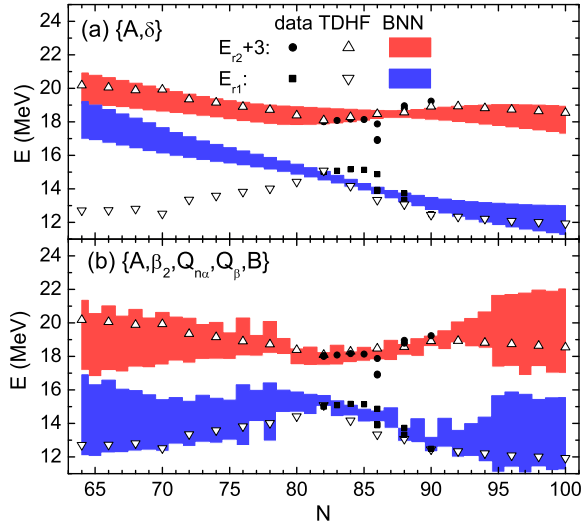}
	\caption{Data of GDR energies in nuclei $^{124-160}$Nd compared to the predictions by BNN using (a) two dimensional input layer $\{A, \delta\}$, and (b) five dimensional input layer $\{A, \beta_{2}, Q_{n\alpha}, Q_{\beta}, B \}$. In order to distinguish high GDR energies $E_{r2}$ from the low one $E_{r1}$, the data and predictions of high GDR energies are upward moved by 3 MeV. }
	\label{fig-4}
\end{figure}

Figure \ref{fig-4} shows the low and high GDR energies in $^{124-160}$Nd as a function of neutron number.
The Nd isotopes display the typical evolution of the quadrupole deformation \cite{Wang2014}.
Because of the magic number N = 82, the nucleus $^{142}$Nd has a spherical shape.
The quadrupole deformation parameter $\beta_{2}$ as a function of neutron number N shows that the $^{142}$Nd nucleus has a minimum value 0.07, with gradual raises in two sides.
The correlation between the splitting $E_{r2}-E_{r1}$ and the deformation parameter $\beta_{2}$ has been found by the TDHF model in Ref. \cite{Mennana2020}.
The TDHF calculations (up and down triangles in Fig. \ref{fig-4}) show that two GDR energies are the same for $^{142}$Nd, but gradually split in both sides of N = 82.
The BNN predictions using input $\{A, \delta\}$ [Fig. \ref{fig-4}(a)] agree to the TDHF calculations for $E_{r2}$, but show a huge difference from them for $E_{r1}$.
After the optimal ground-state properties $\{A, \beta_{2}, Q_{n\alpha}, Q_{\beta}, B \}$ are employed as the input layer to train the BNN, the 95\% confidence intervals of the predictions [Fig. \ref{fig-4}(b)] cover the TDHF calculations, except for the cases of nuclei $^{140,144,146}$Nd.
The selected input layer provides physics guides in the BNN and hence actively contributes to reduce the risk of the non-physics divergence.
For the data (circles and squares in Fig. \ref{fig-4}), both the $^{142}$Nd nucleus and the $^{144,145,146}$Nd nuclei have the same low and high GDR energies.
This is different from the TDHF calculations.
In fact, the GDR energies are extracted from the $^{144,145,146}$Nd($\gamma$,sn) spectra.
The $^{144,145,146}$Nd($\gamma$,sn) spectra (Graph 102-105 in Ref. \cite{Plujko2018}) display asymmetrical shapes, but were fitted by single peak function.
Both the TDHF calculations and BNN predictions indicate that two peaks fitting to the GDR spectra is necessary for $^{144,145,146}$Nd nuclei.
It should be interesting to train the BNN by the GDR spectra by not the GDR energies, which will be our future efforts.

In conclusion, it has been proved in previous works \cite{Utama2016, Utama2016a, Neufcourt2018, Wang2019, Ma2020} that one can include as much physics as possible in the initial prediction by
using the physics motivated models and use the BNN to fine tune these models by modeling the residuals.
In this work, the case study on prediction of giant dipole resonance (GDR) energy is presented to illustrate the new method to provide physics guides in BNN from input layer without the initial prediction by the physics motivated model.
The microscopic theory such as Hartree-Fock-based random phase approximation and time-dependent Hartree-Fock theory has proved that the GDR energies depend strongly on the ground-state wave function \cite{Bonasera2018, Mennana2020}.
Based on this dependence, it is proposed to predict the GDR energies by using the data of the nuclear properties in the ground state, which are abundant and credible.
The Spearman's correlation coefficients are applied to assess the statistical dependence between the ground-state properties and the GDR energies.
Then the optimal ground-state properties are selected as neurons of the input layer in the BNN for prediction of the GDR energies.
It is shown that those selected ground-state properties provide physics guides in the BNN and hence actively contributes to reduce the risk of the non-physics divergence.
This work gives a demonstration to find effects of the GDR energy by using the BNN without the physics
motivated model, which may be helpful for discovering physics effects from the complex nuclear data.

\section*{ACKNOWLEDGMENTS}

This work was supported by the National Natural Science Foundation of China under Grants Nos. 11875328 and 12075327.

\bibliographystyle{IEEEtran}
\bibliography{mybibfile}

\begin{thebibliography}{10}
\providecommand{\url}[1]{#1}
\csname url@samestyle\endcsname
\providecommand{\newblock}{\relax}
\providecommand{\bibinfo}[2]{#2}
\providecommand{\BIBentrySTDinterwordspacing}{\spaceskip=0pt\relax}
\providecommand{\BIBentryALTinterwordstretchfactor}{4}
\providecommand{\BIBentryALTinterwordspacing}{\spaceskip=\fontdimen2\font plus
\BIBentryALTinterwordstretchfactor\fontdimen3\font minus
  \fontdimen4\font\relax}
\providecommand{\BIBforeignlanguage}[2]{{%
\expandafter\ifx\csname l@#1\endcsname\relax
\typeout{** WARNING: IEEEtran.bst: No hyphenation pattern has been}%
\typeout{** loaded for the language `#1'. Using the pattern for}%
\typeout{** the default language instead.}%
\else
\language=\csname l@#1\endcsname
\fi
#2}}
\providecommand{\BIBdecl}{\relax}
\BIBdecl

\bibitem{Gernoth1993}
K.~Gernoth, J.~Clark, J.~Prater, and H.~Bohr, ``Neural network models of
  nuclear systematics,'' \emph{Physics Letters B}, vol. 300, no. 1-2, pp. 1--7,
  feb 1993.

\bibitem{Athanassopoulos2004}
S.~Athanassopoulos, E.~Mavrommatis, K.~Gernoth, and J.~Clark, ``Nuclear mass
  systematics using neural networks,'' \emph{Nuclear Physics A}, vol. 743,
  no.~4, pp. 222--235, nov 2004.

\bibitem{Costiris2009}
N.~J. Costiris, E.~Mavrommatis, K.~A. Gernoth, and J.~W. Clark, ``Decoding
  beta-decay systematics: A global statistical model for beta-half-lives,''
  \emph{Physical Review C}, vol.~80, no.~4, p. 044332, oct 2009.

\bibitem{Bayram2014}
T.~Bayram, S.~Akkoyun, and S.~O. Kara, ``A study on ground-state energies of
  nuclei by using neural networks,'' \emph{Annals of Nuclear Energy}, vol.~63,
  pp. 172--175, jan 2014.

\bibitem{Utama2016}
R.~Utama, J.~Piekarewicz, and H.~B. Prosper, ``Nuclear mass predictions for the
  crustal composition of neutron stars: A bayesian neural network approach,''
  \emph{Physical Review C}, vol.~93, no.~1, p. 014311, jan 2016.

\bibitem{Utama2016a}
R.~Utama, W.-C. Chen, and J.~Piekarewicz, ``Nuclear charge radii: density
  functional theory meets bayesian neural networks,'' \emph{Journal of Physics
  G: Nuclear and Particle Physics}, vol.~43, no.~11, p. 114002, oct 2016.

\bibitem{Neufcourt2018}
L.~Neufcourt, Y.~Cao, W.~Nazarewicz, and F.~Viens, ``Bayesian approach to
  model-based extrapolation of nuclear observables,'' \emph{Physical Review C},
  vol.~98, no.~3, p. 034318, sep 2018.

\bibitem{Wang2019}
Z.-A. Wang, J.~Pei, Y.~Liu, and Y.~Qiang, ``Bayesian evaluation of incomplete
  fission yields,'' \emph{Physical Review Letters}, vol. 123, no.~12, p.
  122501, sep 2019.

\bibitem{Ma2020}
C.-W. Ma, D.~Peng, H.-L. Wei, Z.-M. Niu, Y.-T. Wang, and R.~Wada, ``Isotopic
  cross-sections in proton induced spallation reactions based on the bayesian
  neural network method,'' \emph{Chinese Physics C}, vol.~44, no.~1, p. 014104,
  jan 2020.

\bibitem{Niu2018}
Z.~Niu and H.~Liang, ``Nuclear mass predictions based on bayesian neural
  network approach with pairing and shell effects,'' \emph{Physics Letters B},
  vol. 778, pp. 48--53, mar 2018.

\bibitem{Baldwin1947}
G.~C. Baldwin and G.~S. Klaiber, ``Photo-fission in heavy elements,''
  \emph{Physical Review}, vol.~71, no.~1, pp. 3--10, jan 1947.

\bibitem{Ishkhanov2007}
B.~S. Ishkhanov and V.~N. Orlin, ``Semimicroscopic description of the giant
  dipole resonance,'' \emph{Physics of Particles and Nuclei}, vol.~38, no.~2,
  pp. 232--254, mar 2007.

\bibitem{Bracco2019}
A.~Bracco, E.~Lanza, and A.~Tamii, ``Isoscalar and isovector dipole
  excitations: Nuclear properties from low-lying states and from the isovector
  giant dipole resonance,'' \emph{Progress in Particle and Nuclear Physics},
  vol. 106, pp. 360--433, may 2019.

\bibitem{Ceruti2017}
S.~Ceruti, F.~Camera, A.~Bracco, A.~Mentana, R.~Avigo, G.~Benzoni, N.~Blasi,
  G.~Bocchi, S.~Bottoni, S.~Brambilla, F.~C.~L. Crespi, A.~Giaz, S.~Leoni,
  B.~Million, A.~I. Morales, R.~Nicolini, L.~Pellegri, S.~Riboldi, O.~Wieland,
  D.~Bazzacco, M.~Ciemala, E.~Farnea, A.~Gottardo, M.~Kmiecik, A.~Maj,
  D.~Mengoni, C.~Michelagnoli, V.~Modamio, D.~Montanari, D.~Napoli, F.~Recchia,
  E.~Sahin, C.~Ur, J.~J. Valiente-Dob{\'{o}}n, B.~Wasilewska, and
  M.~Zieblinski, ``Experimental study of the isovector giant dipole resonance
  in zr80 and rb81,'' \emph{Physical Review C}, vol.~95, no.~1, p. 014312, jan
  2017.

\bibitem{Mondal2018}
D.~Mondal, D.~Pandit, S.~Mukhopadhyay, S.~Pal, S.~Bhattacharya, A.~De, N.~D.
  Dang, N.~Q. Hung, S.~Bhattacharya, S.~Bhattacharyya, B.~Dey, P.~Roy,
  K.~Banerjee, and S.~Banerjee, ``Study of giant dipole resonance in hot
  rotating light mass nucleus 31p,'' \emph{Physics Letters B}, vol. 784, pp.
  423--428, sep 2018.

\bibitem{Kobayashi2019}
N.~Kobayashi, K.~Miki, T.~Hashimoto, C.~Iwamoto, A.~Tamii, N.~Aoi, M.~P.
  Carpenter, K.~Hatanaka, J.~Isaak, E.~Ideguchi, S.~Morinobu, S.~Nakamura, and
  S.~Noji, ``Excitation and $\gamma$-decay coincidence measurements at the graf
  beamline for studies of pygmy and giant dipole resonances,'' \emph{The
  European Physical Journal A}, vol.~55, no.~12, dec 2019.

\bibitem{Tian2019}
Y.~Tian, X.~Tao, J.~Wang, X.~Ke, R.~Xu, and Z.~Ge, ``Giant dipole resonance
  parameters from photoabsorption cross-sections,'' \emph{Chinese Physics C},
  vol.~43, no.~11, p. 114102, oct 2019.

\bibitem{Plujko2018}
V.~Plujko, O.~Gorbachenko, R.~Capote, and P.~Dimitriou, ``Giant dipole
  resonance parameters of ground-state photoabsorption: Experimental values
  with uncertainties,'' \emph{Atomic Data and Nuclear Data Tables}, vol.
  123-124, pp. 1--85, sep 2018.

\bibitem{Lyutorovich2015}
N.~Lyutorovich, V.~Tselyaev, J.~Speth, S.~Krewald, F.~Grümmer, and P.-G.
  Reinhard, ``Isoscalar and isovector giant resonances in a self-consistent
  phonon coupling approach,'' \emph{Physics Letters B}, vol. 749, pp. 292--297,
  oct 2015.

\bibitem{Bonasera2018}
G.~Bonasera, M.~R. Anders, and S.~Shlomo, ``Giant resonances in
  ca40,48,~ni68,~zr90,~sn116,~sm144 , and pb208,'' \emph{Physical Review C},
  vol.~98, no.~5, p. 054316, nov 2018.

\bibitem{Fracasso2012}
S.~Fracasso, E.~B. Suckling, and P.~D. Stevenson, ``Unrestricted skyrme-tensor
  time-dependent hartree-fock model and its application to the nuclear response
  from spherical to triaxial nuclei,'' \emph{Physical Review C}, vol.~86,
  no.~4, p. 044303, oct 2012.

\bibitem{Mennana2020}
A.~A.~B. Mennana, Y.~E. Bassem, and M.~Oulne, ``Giant dipole resonance and
  shape evolution in nd isotopes within {TDHF} method,'' \emph{Physica
  Scripta}, vol.~95, no.~6, p. 065301, mar 2020.

\bibitem{Wang2020}
R.~Wang, Z.~Zhang, L.-W. Chen, C.~M. Ko, and Y.-G. Ma, ``Constraining the
  in-medium nucleon-nucleon cross section from the width of nuclear giant
  dipole resonance,'' \emph{Physics Letters B}, vol. 807, p. 135532, aug 2020.

\bibitem{Kong2017}
H.-Y. Kong, J.~Xu, L.-W. Chen, B.-A. Li, and Y.-G. Ma, ``Constraining
  simultaneously nuclear symmetry energy and neutron-proton effective mass
  splitting with nucleus giant resonances using a dynamical approach,''
  \emph{Physical Review C}, vol.~95, no.~3, p. 034324, mar 2017.

\bibitem{He2014}
W.~He, Y.~Ma, X.~Cao, X.~Cai, and G.~Zhang, ``Giant dipole resonance as a
  fingerprint of $\alpha$ clustering configurations {inC}12ando16,''
  \emph{Physical Review Letters}, vol. 113, no.~3, p. 032506, jul 2014.

\bibitem{Su2019}
J.~Su, ``Constraining symmetry energy at subnormal density by isovector giant
  dipole resonances of spherical nuclei,'' \emph{Chinese Physics C}, vol.~43,
  no.~6, p. 064109, may 2019.

\bibitem{Su2020}
J.~Su, L.~Zhu, and C.~Guo, ``Constraints on the effective mass splitting by the
  isoscalar giant quadrupole resonance,'' \emph{Physical Review C}, vol. 101,
  no.~4, p. 044606, apr 2020.

\bibitem{Colo2014}
G.~Col{\`{o}}, U.~Garg, and H.~Sagawa, ``Symmetry energy from the nuclear
  collective motion: constraints from dipole, quadrupole, monopole and
  spin-dipole resonances,'' \emph{The European Physical Journal A}, vol.~50,
  no.~2, feb 2014.

\bibitem{Chakrabarty2016}
D.~R. Chakrabarty, N.~D. Dang, and V.~M. Datar, ``Giant dipole resonance in hot
  rotating nuclei,'' \emph{The European Physical Journal A}, vol.~52, no.~5,
  may 2016.

\bibitem{Pandit2019}
D.~Pandit, S.~Bhattacharya, D.~Mondal, B.~Dey, S.~Mukhopadhyay, S.~Pal, A.~De,
  and S.~R. Banerjee, ``Role of fluctuations in a thermal phase transition in a
  nucleus probed via the giant dipole resonance,'' \emph{Physical Review C},
  vol.~99, no.~2, p. 024315, feb 2019.

\bibitem{Huang2017}
W.~Huang, G.~Audi, M.~Wang, F.~G. Kondev, S.~Naimi, and X.~Xu, ``The {AME}2016
  atomic mass evaluation (i). evaluation of input data; and adjustment
  procedures,'' \emph{Chinese Physics C}, vol.~41, no.~3, p. 030002, mar 2017.

\bibitem{Audi2017}
G.~Audi, F.~G. Kondev, M.~Wang, W.~Huang, and S.~Naimi, ``The {NUBASE}2016
  evaluation of nuclear properties,'' \emph{Chinese Physics C}, vol.~41, no.~3,
  p. 030001, mar 2017.

\bibitem{Wang2017}
M.~Wang, G.~Audi, F.~G. Kondev, W.~Huang, S.~Naimi, and X.~Xu, ``The {AME}2016
  atomic mass evaluation ({II}). tables, graphs and references,'' \emph{Chinese
  Physics C}, vol.~41, no.~3, p. 030003, mar 2017.

\bibitem{Wang2014}
N.~Wang, M.~Liu, X.~Wu, and J.~Meng, ``Surface diffuseness correction in global
  mass formula,'' \emph{Physics Letters B}, vol. 734, pp. 215--219, jun 2014.

\bibitem{MacKay1992}
D.~J.~C. MacKay, ``A practical bayesian framework for backpropagation
  networks,'' \emph{Neural Computation}, vol.~4, no.~3, pp. 448--472, may 1992.

\bibitem{bishop1995neural}
C.~M. Bishop, \emph{Neural networks for pattern recognition}.\hskip 1em plus
  0.5em minus 0.4em\relax Oxford university press, 1995.

\bibitem{neal2012bayesian}
R.~M. Neal, \emph{Bayesian learning for neural networks}.\hskip 1em plus 0.5em
  minus 0.4em\relax Springer Science \& Business Media, 2012, vol. 118.

\bibitem{Mullachery2018}
V.~Mullachery, A.~Khera, and A.~Husain, ``Bayesian neural networks,''
  \emph{arXiv preprint arXiv:1801.07710}, 2018.

\bibitem{salimans2015markov}
T.~Salimans, D.~Kingma, and M.~Welling, ``Markov chain monte carlo and
  variational inference: Bridging the gap,'' in \emph{International Conference
  on Machine Learning}, 2015, pp. 1218--1226.

\bibitem{Blundell2015}
C.~Blundell, J.~Cornebise, K.~Kavukcuoglu, and D.~Wierstra, ``Weight
  uncertainty in neural networks,'' \emph{arXiv preprint arXiv:1505.05424},
  2015.

\end{thebibliography}

\end{document}